\documentclass[aps,prl,twocolumn,floatfix,showpacs,showkeys,amssymb]{revtex4}

\usepackage{graphicx}

\usepackage{amsmath}

\usepackage{amsfonts}

\usepackage{amssymb}

\begin{document}
\topmargin -15mm

\def\ba{\begin{array}}

\def\ea{\end{array}}

\def\be{\begin{eqnarray}}

\def\ee{\end{eqnarray}}

\def\bea{\begin{equation}\begin{array}{l}}

\def\eea{\end{array}\end{equation}}

\def\f#1#2{\frac{\displaystyle #1}{\displaystyle #2}}

\def\om{\omega}

\def\Om{\Omega}

\def\omm{\omega^a_b}

\def\we{\wedge}

\def\de{\delta}

\def\De{\Delta}

\def\va{\varepsilon}

\def\omb{\bar{\omega}}

\def\la{\lambda}

\def\vv{\f{V}{\la^d}}

\def\si{\sigma}

\def\t{T_+}

\def\v{v_{cl}}

\def\m{m_{cl}}

\def\n{N_{cl}}

\def\bi{\bibitem}

\def\c{\cite}

\def\sa{\sigma_{\alpha}}

\def\ua{\uparrow}

\def\da{\downarrow}

\def\mua{\mu_{\alpha}}

\def\ga{\gamma_{\alpha}}

\def\g{\gamma}

\def\ora{\overrightarrow}

\def\pa{\partial}

\def\ov{\ora{v}}

\def\al{\alpha}

\def\bt{\beta}

\def\R{R_{\rm eff}}

\def\muu{\f{\mu}{ed}}

\def\E{\f{edE(\tau)}{\om}}

\def\t{\tau}

\title{Quasienergy spectra of a charged particle in planar honeycomb lattices } 

\author{Wei Zhang $^1$
{\footnote {Email address: zhang $\_$wei@iapcm.ac.cn}}, Ping Zhang $^{1,2}$,
Suqing Duan $^1$, and Xian-geng Zhao $^1$}
\affiliation{$^1$Institute of Applied Physics and Computational
Mathematics, P.O.Box 8009, Beijing 100088, P. R. China\\
$^2$Center for Applied Physics and Technology, Peking University, Beijing 100871, P. R. China}
\date{\today } 

\begin{abstract}
The low energy spectrum of a particle in planar honeycomb lattices
is conical, which leads to the unusual electronic properties of
graphene. In this letter we calculate the quasienergy spectra of a
charged particle in honeycomb lattices driven by a strong AC field,
which is of fundamental importance for its time-dependent dynamics.
We find that depending on the amplitude, direction and frequency of
external field, many interesting phenomena may occur, including band
collapse, renormalization of velocity of ``light'', gap opening
etc.. Under suitable conditions, with increasing the magnitude of
the AC field, a series of phase transitions from gapless phases to
gapped phases appear alternatively. At the same time, the Dirac
points may disappear or change to a line. We suggest  possible
realization of the system in Honeycomb optical lattices.


\end{abstract}



\maketitle




Lattice structure has important impact on the dynamics of a system
through the related band structure. For many materials with
square/cubic lattice, their energy spectra near the bottom of
conduction band are parabolic and the low energy excitation is a
quasiparticle with an effective mass. However graphene, a
two-dimensional monolayer of carbon atoms, has a linear dispersion
relation near so-called Dirac points due to its special honeycomb
lattice\c{graenergy}. Therefore in the long distance limit, the
basic excitation is a  massless Dirac Fermion. There are extensive
ongoing experimental and theoretical studies on graphene
\c{graphene}. On one hand graphene provides an excellent
condensed-matter analogue of (2+1) dimensional quantum
electrodynamics, where the effective velocity of ``light'' is  1/300
of the real velocity of light.  On the other hand the unusual energy
spectrum also leads to novel transport properties, such as anomalous
quantum Hall effect \c{qhe}. Besides graphene, the optical honeycomb
lattice has also been realized in experiments \c{oplhoney}. The
flexible tunability of optical lattices gives people much more
opportunities to explore the interesting physics in two-dimensional
systems with honeycomb lattices. Recently much attention has been
drawn to this direction, for instance  Haldane's quantum Hall
effects was proposed in ultracold atoms in optical honeycomb lattice
\c{xing}, P-orbital physics in the honeycomb optical lattice  was
also addressed \c{dassarma}, Bose-Einstein condensation in a
honeycomb optical lattice was considered in \c{bechopl}, just name a
few.

An important issue here is the adjustment of band structure. One
convenient and effective way is through external time-dependent
field. For a system driven by a time-periodic electric field, band
suppression \c{bandsupp} and band collapse\c{bandcoll} appear. There
are also many other interesting phenomena, including coherent
destruction of tunneling\c{cdt},
 dynamical
localization\c{dynlocal}, photovoltalic effects\c{phovol}, and
photon-assistant Fano resonance \c{photfano} etc.. Many of them have
been observed in electronic system and/or optical lattices. As the
energy spectrum is essential for studying the optical and electronic
properties of the relevant system, the quasienergy spectrum plays a
key role in understanding the time-dependent phenomena.

Thus it is of great importance and interest to  study the
quasienergy spectra of a particle in honeycomb lattice in the
presence of an AC field. This issue is addressed in detail in this
paper. We show that the quasienergy spectrum of honeycomb lattice
has a quite rich structure. A series of phase transitions (from
gapless phase to gapped phase) appear with changing of the
amplitude, direction, and frequency of the external electric field,
which is quite easy  in experiments. In particular,
 under special conditions the
particle can be localized in one direction, or two directions, thus
the dimension of the system reduces effectively from two dimension (2D) to one
dimension (1D) and to zero dimension (0D). The changing of external
field may also lead to the effective mass generation and
renormalization of the velocity of ``light''.
\begin{figure}[tbh]
\includegraphics*[width=0.45\linewidth,angle=270]{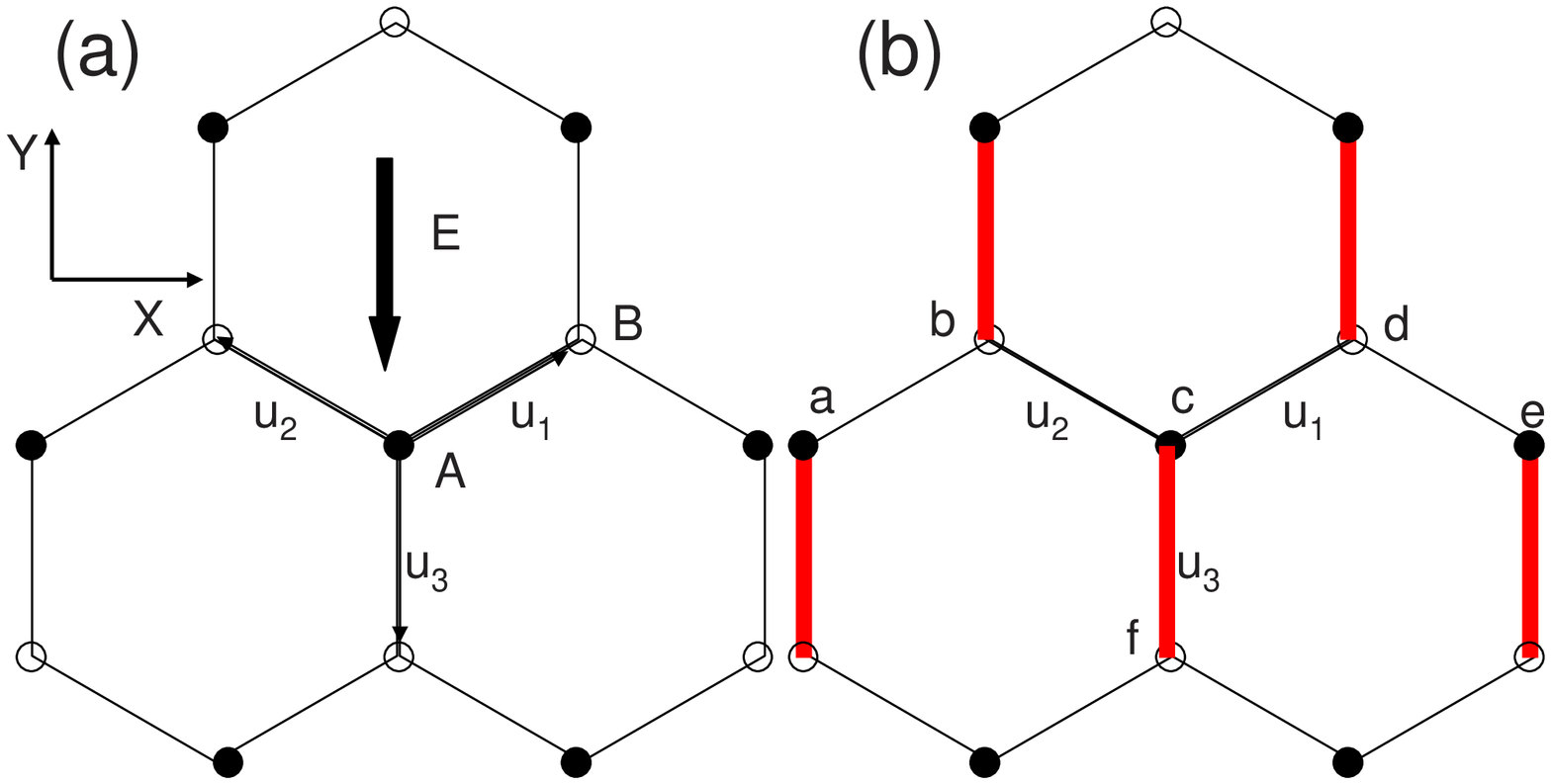}
\caption{(a) Planar honeycomb lattice with electric field along Y
direction. Black dots are sites of sublattice A and white dots are
sites of sublattice B. (b) The effective dimerization bond pattern.
The effective hopping constants along the (red) thick bonds are
different from those along (black) thin bonds.
 } \label{fig1}
\end{figure}

 The planar honeycomb lattice (Fig.1(a)) consists of two equivalent sublattices A (black dots) and B (white dots).
We consider the charged particle hopping between the nearest
neighbor
 sites from different sublattices. In addition, the system under consideration is
subject to a time-periodic electric field ${\bf E}(t)={\bf E}_0
\cos (\om t) $.
The tight-binding Hamiltonian has the form \be
H=\sum_{<i,j>} V (c_i^+ c_j+c_j^+ c_i)+\sum_i e{\bf E} (t) \cdot{\bf
r}_i c_i^+c_i,
 \ee
where $<i,j>$ refers to the nearest neighbor lattice sites, ${\bf
r}_i$ is the position of site $i$. The probability amplitude $\al
({\bf r}_i)$ for a particle at site $i$ of sublattice A (or $\beta
({\bf r}_i)$ at site of sublattice B ) can be calculated by solving
the corresponding Schr$\ddot{o}$dinger equation. It is very helpful
to make the unitary transformation $\al({\bf r}_i) \rightarrow
\al({\bf r}_i) e^{-i {\bf A}(t)\cdot {\bf r}_i}$ ($\beta ({\bf r}_i)
\rightarrow \beta ({\bf r}_i) e^{i {\bf A}(t)\cdot {\bf r}_i}$),
where ${\bf A}(t)=\f{e {\bf E}_0 a}{\om} \sin(\om t)$, with $a$ the
lattice constant. The subsequent
 Fourier transformation leads to
 \bea
 i \f{d}{dt}\left( \ba{c}
\al({\bf k})\\
\beta({\bf k})\ea \right) \,

=H({\bf k},t) \left( \ba{c}
\al({\bf k})\\
\beta({\bf k})\ea \right) \, ,\eea

\bea

H= \left( \ba{cc}
0 &  V\sum_{j=1,2,3} e^{i(a{\bf k}-{\bf A}(t))\cdot {\bf u}_j}\\
 V\sum_{j=1,2,3} e^{-i(a{\bf k}-{\bf A}(t))\cdot {\bf u}_j} & 0 \ea \right)
 \,
 \eea

Due to the Floquet theorem, the quasienergy can be obtained by
diagonalizing the time evolution operator $\hat{T} \int_0^T   dt
e^{-i H({\bf k},t)}$, where $\hat{T}$ refers to time ordering and T
is the period of the AC field . We first consider the case of
electric field along Y direction. In the high frequency limit $V/\om
<<1$, we can obtain the following analytical result for quasienergy
spectrum \be \va=\pm
V\sqrt{A^2+4B^2(\cos\f{\sqrt{3}}{2}k_xa)^2+4AB\cos\f{\sqrt{3}}{2}k_xa\cos\f{3}{2}k_ya},\ee
where $A=J_0 (eE_0a/\om)$ and $B=J_0 (eE_0a/2\om)$, $J_0$ is the
zero-th order Bessel function. When $E_0 \rightarrow 0$, $A,B
\rightarrow 1$, we reproduce  the results for energy spectrum of a
system with honeycomb lattice without AC electric field.



{\bf Quantum phase transition}
The conduction band and valence band may form conically shaped
valleys that touch at some specific points (called conical points or
Dirac points). These Dirac points are determined as $k_ya=2\pi/3$,
$\cos(\f{\sqrt{3}}{2}k_xa)=A/2B$. It is clear that there is no
solution, when $|A|>2|B|$. In this situation, an energy gap is
generated. In Fig. 2 (red solid curve) we show
$|A|/2|B|=|J_0(x)/2J_0(x/2)|$ versus $x=eE_0a/\om$.
\begin{figure}[tbh]
\includegraphics*[width=0.75\linewidth]{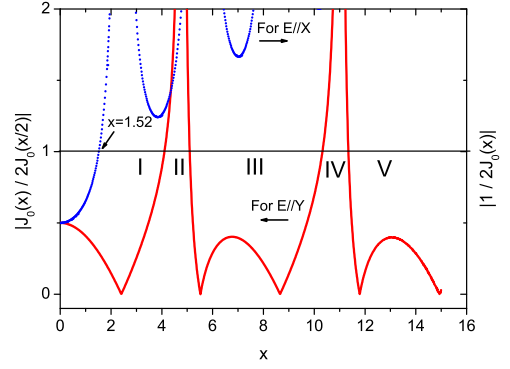}
\caption{Red solid curve: $|A|/2|B|=|J_0(x)/2J_0(x/2)|$ versus x;
Blue dotted curve: $|1/2J_0(x)|$ versus x.} \label{fig3}
\end{figure}
From Fig. 2 it can be seen that gap opening happens in the regimes (for $eE_0a/\om$)
II=[4.1, 5.1], IV=[10.3, 11.3],....
Figures 3(a)-(f) (corresponding to regimes I to V) show the
quasienergy spectra around the minimum of conduction band/maximun of
valence band for different values of $eE_0a/\om$, which are obtained
by exact numerical calculation based on Eqs. (2) and (3).

\begin{figure}[tbh]
\includegraphics*[width=1.0\linewidth]{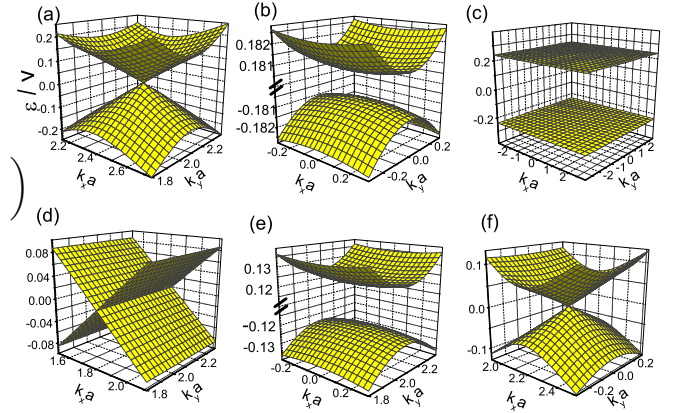}
\caption{Quasienergy spectra for $eE_0a/\om$=3.5 (in regime I),
$eE_0a/\om$=4.60 (in regime II), $eE_0a/\om$=4.81 (band collapse
condition), $eE_0a/\om$=5.52 (in regime III),
 $eE_0a/\om$=10.8 (in regime IV), and $eE_0a/\om$=12.5
(in regime V), are obtained by exact numerical calculation.
$V/\om=0.02$. } \label{fig5ab}
\end{figure}



We can see from figure 3 that as $eE_0a/\om$ increases, the system
undergoes a series of  phases corresponding to the regimes I-V,
i.e., gapless phase (regime I) $\rightarrow$ gapped phase (regime
II) $\rightarrow$ gapless phase (regime III) $\rightarrow$ gapped
phase (regime IV) $\rightarrow$ gapless phase (regime V). The gap
opening introduces a new energy scale and may also lead to the
generation of an effective mass as seen in Figs. 3(b) and (e), since
the quasienergy spectrum near the bottom of conduction band is
parabolic. We find the gap and the effective mass are proportional
to $D=(|A|-2|B|) \theta(|A|-2|B|)$. Thus D is the order parameter,
which is zero in the massless phases and nonzero in the gapped
phases. As one can see from figure 3, with tuning AC field, the
Dirac points may disappear in the gapped phase, or change from
discrete points to a line as shown in Fig. 3(d). Near $\va=0$, the
density of quasienergy states $D(\va)$ is zero for gapped phases,
$D(\va)\sim |\va|$ near discrete Dirac points, and $D(\va) \sim$
constant near the line. According to the generalized Kubo-Greenwood
formula for time-dependent system\c{shi}, the conductivity depends
on the density of quasienergy states. Therefore by simply tuning the
strength of external AC field, we can realize the transitions from
``metal'' phase to  ``semimetal'' phase and to
``semiconductor''/``insulator'' phase.

{\bf Band collapse} A typical character of  Fig. 2 is that
$|A|/2|B|$ is divergent at $eE_0a/\om=4.81, 11.0,...$. By checking
Eq. (4), it is easy to see that band collapse happens and a gap
opens under this condition B=0 or $J_0 (eE_0a/2\om)=0$.
Figure 3(c) shows the quasienergy spectrum for $eE_0a/\om=4.81$,
twice of 2.405, the first zero point of $J_0$.
As expected, we see the flat conduction and valence bands, which are
drastically different from the remaining figures of Fig. 3, and the
case without an AC field. Moreover we see a gap between conduction
and valence bands opens. From eq. (4), one can also see that the
quasienergy band collapses in the Y direction, when A=0, i.e.,
$eE_0a/\om=2.405$,....

{\bf Time-dependent dynamics } The collapse of  quasienergy spectrum
implies  dynamical localization. To verify this picture, we
calculate the time evolution of a particle initially at the origin.
In our calculation, the system size is 101 $\times$ 101, time
step=0.02 (in unit of $1/\om$), and the hopping constant is
$V/\om=0.02$ as that in the calculation of quasienergy. In the long
time regime, the mean square displacement of the particle shows the
time-dependent behavior as $t^\al$. In figure 4(a) we show the time
dependence of the mean square displacement. We can see that when the
system is  away from band collapse condition ($eE_0a/\om=4.81$),
$\al=2$, the dynamics is a typical ballistic motion.
Under the band collapse condition, the mean quare displacement is
finite ($<1$) for all the time, i.e., $\al=0$, indicating the
dynamical localization of the particle due to the presence of an AC
field. In Fig. 4(b), we show  the time dependence of the mean square
displacement in the X and Y directions for $eE_0a/\om=2.405$, under
which the quasienergy spectrum is flat in the Y direction. It is
clear that the particle is localized in the Y direction ($<m^2_y>
<0.18$), indicating a dimension reduction from 2D to 1D.

\begin{figure}[tbh]
\includegraphics*[width=0.9\linewidth]{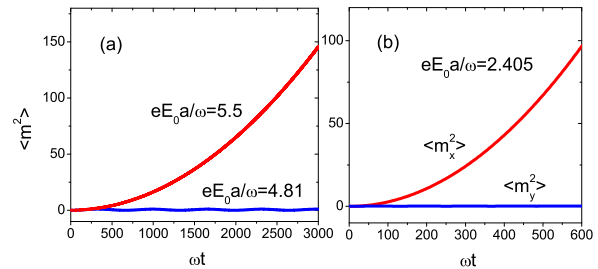}
\caption{(a) Time dependence of the mean square-displacement for
$eE_0a/\om=4.81$ and $eE_0a/\om=5.50$; (b)Time dependence of the mean
square-displacement in the X and Y directions for $eE_0a/\om=2.405$.}
\label{fig6}
\end{figure}


{\bf Renormalization of velocity of ``light''} For system in the
gapless phases, we may  obtain the velocity of ``light'' by
expanding  the quasienergy spectrum around Dirac points
 \bea
v_x=2Bt\sqrt{3}\sin(\f{\sqrt{3}}{2}\bar{k}_xa)\\
v_y=6Bt\cos(\f{\sqrt{3}}{2}\bar{k}_xa), \eea where $\{\bar{k}_xa,
\bar{k}_ya\}$ refer to the Dirac points. We can see that the
velocity of ``light'' can be tuned by the amplitude and frequency of
external electric field. From Fig. 5(a), one can see that the
velocity of ``light'' for $eE_0a/\om=3.5$ reduces to  $ \sim
\f{1}{3}$ of that without external electric field.
The external field may also lead to the anisotropy of velocity of
``light'', see figure 5(b) and Eq. (5). An extremal example is that
the velocity in one direction is finite, while in the other
direction becomes zero under the condition A=0, i.e.
$\cos(\f{\sqrt{3}}{2}\bar{k}_xa)=0$, as shown in Fig. 3(d).
The tunability of velocity of ``light'' gives us
more opportunities to explore relativistic effects in electronic
systems or optical lattices.






\begin{figure}[tbh]
\includegraphics*[width=1.0\linewidth]{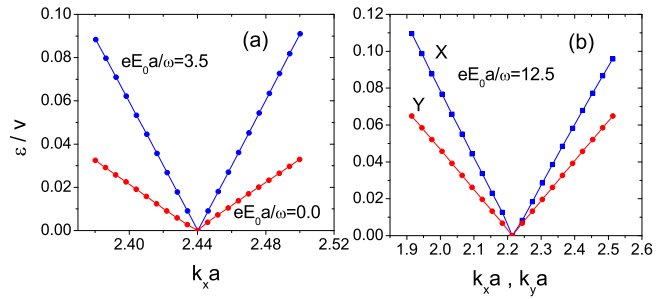}
\caption{ Renormalization of the velocity of ``light''. (a)
Quasienergy spectrum for $eE_0a/\om$=0.0 and 3.5. (b) Anisotropy of
velocity of ``light''. Quasienergy spectrum for $eE_0a/\om=12.5$.
For the purpose of comparison, we have shifted the coordinates of
Dirac points. } \label{fig7}
\end{figure}




{\bf Dependence on the direction of electric fields} The quasienergy
spectrum is dependent on the direction of electric field. For the
case of an electric field along X direction, the quasienergy
spectrum still takes the form of Eq. (4) with A=1 and
$B=J_0(\f{\sqrt{3}}{2}eE_0a/\om)$. In this case the condition for
the existence of Dirac points is
$|2J_0(\f{\sqrt{3}}{2}eE_0a/\om)|\geqslant 1$. From Figure 2(b)
(dotted blue curve), we see that when $\f{\sqrt{3}}{2}eE_0a/\om
>1.52$ or $eE_0a/\om >1.76$, a gap always opens, which is quite
contrast to the case with electric field along Y direction discussed
above. The band collapse appears under the new condition
$J_0(\f{\sqrt{3}}{2}eE_0a/\om)=0$.

In the more general situation, the quasienergy spectrum has the form
$\va=\pm V\sqrt{|\sum_j a_j e^{-i a{\bf k}\cdot {\bf u}_j}|^2}$,
where $a_j =J_0(ea{\bf E}_0\cdot{\bf u}_j/\om)$. The general
condition for the existence of Dirac points or the gapless phase is
that
 $|a_j|$ can form a triangle, or $|a_i|+|a_j|\geqslant |a_k|,
\{i,j,k\}=\{1,2,3\}$. The conditions for the band collapse/dynamical
localization become $a_i=a_j=0, i,j \in \{1,2,3\}$.  These
conditions are two equations with two variables, which can be solved
in general situation.

Now we give a simple picture of the phenomena we have found. For an
electric filed ${\bf E}(t)$ in  arbitrary direction, the projection
of this field in three directions along the bond (i.e., directions
${\bf u}_i, i=1,2,3$ (see figure 1)) is $E_i$. Then the effective
hopping constants  along the bonds ${\bf u}_i$ are $V^{eff}_i=V
J_0(eE_i a/\om)$. Thus we may map our system to a system with
honeycomb lattice and effective hopping constants $V^{eff}_i$ along
bond ${\bf u}_i$. The dynamical localization phenomena we have found
are quite clear in this picture. In fact, for the case ${\bf E}//Y$ and
$eE_0a/\om=4.81$, $V^{eff}_1=V^{eff}_2=0$, the particle initially
staying at site c can only oscillate between site c and site f (see
Fig. 1(b)). It is the dynamical localization behavior shown in Fig.
4(a). For the situation $eE_0a/\om=2.405$, $V^{eff}_3=0$, the
particle can only move through the path ``abcde'' shown in Fig.1(b),
which is the dynamical localization in the Y direction show in Fig.
4(b). The occurrence of band gap and effective mass is also a
consequence of the ``bond ordering'', the dimerization pattern (see
Fig. 1(b)). In the continuum limit, the fluctuation of hopping
constant maps to axial vector potential \c{jackiw}. It is different
from the honeycomb lattice with ``Kekule pattern'', in which the
fluctuation of hopping constant maps to the complex-valued Higgs
field in the continuum limit. In that case the opening of energy gap
is related to the spontaneous breaking of an effective axial U(1)
symmetry \c{chamon}, and was proposed for the split of zero-energy
Landau levels in graphene in the presence of magnetic field
\c{hatsugai}. By tuning the external electric field, one may
effectively obtain different bond patterns.

{\bf Realization in optical lattices} The honeycomb optical lattice
can be realized in experiment \c{oplhoney} by using three pairs of
laser beams with co-planar propagating wavevectors $\pm {\bf q}_i,
i=1,2,3$. The magnitude of the three pairs of wavevectors are the
same and their directions form the angle of 120 degree with each
other. The optical potential has the form
 \be
 U=\sum_i U_i=U_0 \sum_i \cos({\bf p}_i \cdot {\bf r}),
 \ee
where ${\bf p}_1={\bf q}_2-{\bf q}_3$, ${\bf p}_2={\bf q}_3-{\bf
q}_1$, and ${\bf p}_3={\bf q}_1-{\bf q}_2$. $U_0$ depends on the
detuning between laser frequency and atomic transition frequency,
and the resonant Rabi frequency proportional to the the quare root
of laser intensity \c{opldyn}. Thus $U_0$ can be tuned quite easily
in experiments. For sufficient large detuning, we may neglect
spontaneous scattering and the high frequency condition $U_0/\om
<<1$ can also be easily met. To simulate the electronic force in
electronic system, we may let the coordinate of the mirror in each
direction ${\bf p}_i$ , reflecting the incoming traveling light
wave, oscillate around its average with $d_i\cos(\om t)$. Its
effect is to replace the atomic effective potential  $U_0 \cos({\bf
p}_i\cdot {\bf r}) $ with $U_0 \cos({\bf p}_i \cdot{\bf r}
-d_i\cos(\om t))$.  In the moving frame of the potential, the atoms
feel an effective force proportional to $d_i \cos(\om t)$, which is
proportional to the electrical field component in the ${\bf p}_i$
direction.

In summary, we have studied the quasienergy spectra and the related
dynamics of a particle in honeycomb lattices driven by an AC field.
By tuning the amplitude, frequency, and direction of the electric
field, we can obtain quite rich phases, including gapped and gapless
phases. Moreover we may tune the velocity of the ``light'', and
realize band collapse and the dimension reduction from 2D to 1D and
0D (dynamical localization in one direction and two directions). In
the current paper, we have focused on the high frequency regime,
i.e. $V/\om <<1$. In graphene with V  around 2.8eV and the lattice
constant $a$ 1.4 {\AA},  our predictions may not be verified
easily. However, they should be quite easy to be observed in optical
lattices. We leave the low frequency regime for future study.

{\bf Acknowledgments} This work was partially supported
  by  the National Science Foundation of China under Grants Nos.
10574017, 10744004 and 10604010.


\end{document}